\documentclass[aps,prd,twocolumn,preprintnumbers,nofootinbib,superscriptaddress,longbibliography]{revtex4-1}

\usepackage[utf8]{inputenc}
\usepackage{amsmath}
\usepackage{amssymb}
\usepackage{xspace}
\usepackage{xcolor}
\usepackage{color}
\usepackage{soul}
\usepackage{url}
\usepackage{appendix}
\usepackage{graphicx}
\usepackage{fancyvrb}
\usepackage{setspace}
\usepackage{siunitx}
\usepackage{braket}
\usepackage{chemformula}
\usepackage[normalem]{ulem}
\usepackage[colorlinks=true,citecolor=blue,linkcolor=blue]{hyperref}

\usepackage[capitalise, english]{cleveref}
\crefformat{equation}{(#2#1#3)} 
\usepackage{bm}

\newcommand*\Bepsilon{\ensuremath{\boldsymbol\epsilon}}
\newcommand*\Bsigma{\ensuremath{\boldsymbol\sigma}}

\newcommand*\Bpartial{\ensuremath{\boldsymbol\partial}}
\renewcommand{\bra}[1]{\left\langle #1\right|}
\renewcommand{\ket}[1]{\left| #1\right\rangle}

\definecolor{red}{rgb}{0.9, 0,0}
\definecolor{cerulean}{rgb}{0., 0.62,0.9}
\definecolor{navy}{rgb}{0.05, 0.05,0.8}

\graphicspath{{./figures/}}

\newcommand{\bfx}{{\bf x}}
\newcommand{\bfD}{{\bf D}}

\newcommand{\bfl}{\ensuremath{\boldsymbol\ell}}
\newcommand{\bfk}{{\bf k}}

\newcommand{\bfJ}{{\bf J}}
\newcommand{\bfS}{{\bf S}}
\newcommand{\bfu}{{\bf u}}

\begin{document}

\title{Broadband phonon production from axion absorption  }

\author{Itay M. Bloch}
\affiliation{Physics Division, Lawrence Berkeley National Laboratory, Berkeley, CA 94720, USA}
\affiliation{Berkeley Center for Theoretical Physics, Department of Physics, University of California, Berkeley, CA 94720, USA}

\author{Simon Knapen}
\affiliation{Physics Division, Lawrence Berkeley National Laboratory, Berkeley, CA 94720, USA}
\affiliation{Berkeley Center for Theoretical Physics, Department of Physics, University of California, Berkeley, CA 94720, USA}

\author{Amalia Madden}
\affiliation{Kavli Institute for Theoretical Physics, Santa Barbara, CA 93106, USA}
\affiliation{Perimeter Institute for Theoretical Physics, Waterloo, Ontario N2L 2Y5, Canada}

\author{Giacomo Marocco}
\affiliation{Physics Division, Lawrence Berkeley National Laboratory, Berkeley, CA 94720, USA}

\begin{abstract}
We show that axion dark matter in the range meV~$\lesssim m_a\lesssim$~100~meV can incoherently excite phonons in crystal targets with unpolarised nuclear spins. This can occur through its coupling to nuclear spins and/or through its induced time-dependent electric dipole moment in nuclei. Due to the random orientation of the nuclear spins, translation symmetry is broken in the phonon effective theory, allowing axion absorption to create phonons with unrestricted momentum. The absorption rate is therefore proportional to the phonon density of states, which generically has support across a wide range of energies, allowing for a broadband detection scheme. We calculate the absorption rate for solid $\text{H}_2$, $\text{D}_2$, $\text{Al}_2\text{O}_3$, $\text{GaAs}$, $\text{H}_2\text{O}$, $\text{D}_2\text{O}$, $\text{Be}$ and $\text{Li}_2 \text{O}$, and find that materials containing light, non-zero spin nuclei are the most promising. The predicted rates for the QCD axion are of the order of a few events / 10 kg-year exposure, setting an ambitious target for the required exposure and background suppression.
\end{abstract}

\maketitle

\section{Introduction}

The nature of Dark Matter (DM) remains an open question in particle physics. Among the many DM candidates, the QCD axion~\cite{Adams:2022pbo} stands out as one of the most well-motivated possibilities. 
Originally proposed to solve the strong CP problem~\cite{Peccei:1977hh, Peccei:1977ur, Wilczek:1977pj, Weinberg:1977ma}, the QCD axion could also account for the observed DM abundance through production mechanisms like the misalignment mechanism~\cite{Preskill:1982cy, Abbott:1982af, Dine:1982ah}. 
Its generalisations, Axion-Like Particles (ALPs), are similarly promising DM candidates, even without a connection to the strong CP problem.\footnote{Since most of the discussion in this paper is identical for ALPs and QCD axions, we will often refer to both as ``axions''. When referring only to the QCD axion, we will make the distinction clear by using the term ``QCD axion''.}

The predictions for the mass of QCD axion DM span many orders of magnitude, depending on the precise production mechanism. 
For post-inflationary axions, the relic abundance may be produced by misalignment for an axion mass $m_a \lesssim \SI{10}{\mu eV}$~\cite{Bae:2008ue,Wantz:2009it,Ballesteros:2016xej, Borsanyi:2016ksw}, while simulations of axions produced by decaying topological string defects point towards the $20$ to $\SI{500}{\mu eV}$ mass range~\cite{Klaer:2017ond, Gorghetto:2020qws, Buschmann:2021sdq}. For modes with long-lived domain walls, their decay suggests values of $m_a$ as large as $1$ to $100$ meV~\cite{Kawasaki:2014sqa,Ringwald:2015dsf, Gorghetto:2020qws, Beyer:2022ywc}. 
This latter mass range, which we focus on in this paper, has proved experimentally challenging to probe in direct detection experiments, although proposals do exist to search for ALPs or the QCD axion~\cite{Horns:2012jf, Baryakhtar:2018doz, BREAD:2021tpx,Mitridate:2020kly,Berlin:2023ppd,Marsh:2022fmo,Schutte-Engel:2021bqm, ARIADNE:2017tdd}, predominantly for the axion-photon coupling. 

In general, the couplings of the QCD axion or an ALP to Standard Model fields depend strongly on the chosen model, and it is possible for the coupling to photons to be suppressed relative to that of quarks~\cite{Craig:2018kne}. For this reason, developing complementary detection strategies that do not rely on the photon coupling is important. 
For QCD axion DM, its defining coupling to gluons generates a time-dependent coupling to the spin of the proton and a time-dependent electric dipole moment for both the proton and the neutron.
Additionally, the QCD axion and ALPs can have derivative couplings to quarks, such as in the Dine-Fischler-Srednicki-Zhitnitsky (DFSZ) model~\cite{Dine:1981rt,Zhitnitsky:1980tq}, which also contribute to their couplings to both protons and neutrons. 

While there are a number of proposals, at this time there is not yet a working experimental setup that is expected to probe a QCD axion in the meV $\lesssim m_a\lesssim$ 100 meV mass range. 
Several existing~\cite{Wu:2019exd, Garcon:2019inh, Bloch:2021vnn, Bloch:2022kjm, Wei:2023rzs, Xu:2023vfn} and proposed experiments~\cite{JacksonKimball:2017elr, Graham:2020kai, Chigusa:2023hmz, Bloch:2024uqb, Arvanitaki:2017nhi} target the ALP coupling to proton and neutron spins; however, they primarily probe a mass range below that considered here.
The 1 meV to 100 meV mass range has proven to be experimentally extremely challenging, due to the shorter DM wavelength and lower number density. 

That said, there are exciting experimental innovations on the horizon in the context of low threshold calorimeters, which aim to ultimately detect individual phonon modes in the 1 meV to 100 meV energy range.  
The TESSERACT program~\cite{PhysRevD.100.092007, PhysRevD.105.092005, HeRALD:2023kww} in particular is currently transitioning from its R\&D phase towards a first full prototype detector, to be housed in the Modane Underground Laboratory.
In this work, we investigate whether a TESSERACT-style experiment could in principle discover the QCD axion or improve on the current astrophysical constraints, which come with uncertainties that can be difficult to quantify.

We will show that an oscillating axion DM background induces a force on the atoms in an ultra-cold crystal, which excites phonons in the crystal. Such individual phonon quanta may then be detectable in single-phonon sensors, such as those being developed within the TESSERACT program.
The rate for the coherent production of phonons through axion-absorption in a spin-polarised crystal was recently calculated ~\cite{Mitridate:2020kly, Mitridate:2023izi}. There, it was shown that the crystal response is resonant at the frequency of the optical phonons at the origin of the Brillouin zone. 
A detection scheme based on this effect is inherently narrow-band, and scanning the optical phonon frequency requires swapping out target materials or applying high amounts of pressure to the crystal~\cite{Ashour:2024xfp}. 
When searching for axion-nucleon couplings, the coherent process is also only present if the spins of the nuclei in the crystal are polarised to a high degree, presenting an additional experimental challenge.

\begin{figure}
    \centering
    \includegraphics[width=\linewidth]{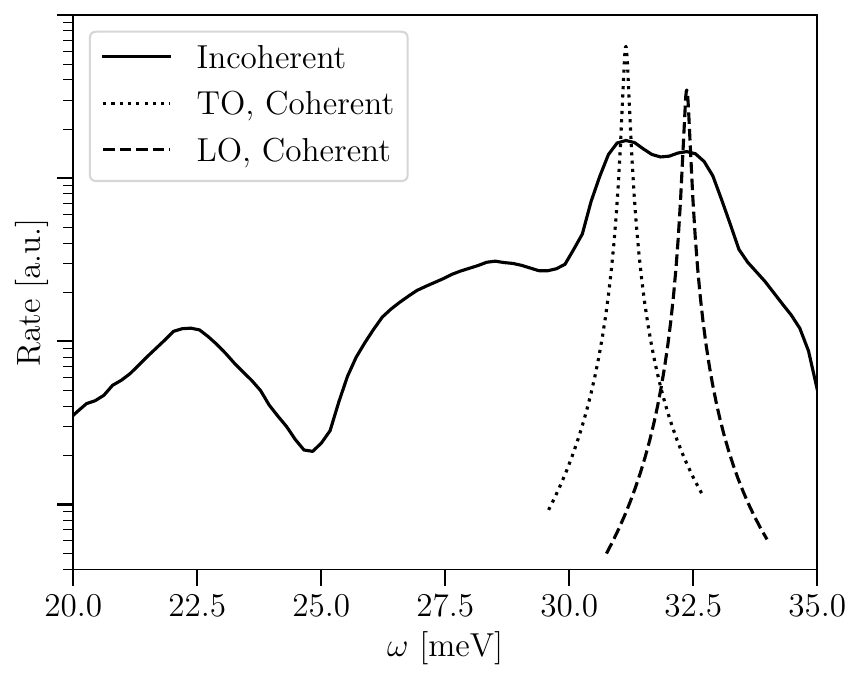}
    \caption{Comparison of rate for coherent and incoherent absorption for GaAs. The phonon lifetime for the longitudinal optical (LO) mode of GaAs is taken from \cite{10.1063/1.358498}. We assume that the lifetime of the transverse optical (TO) modes are of the same order.
    \label{fig:broadbandplot}}
\end{figure}

We calculate the rate at which the axion DM with mass $\SI{1}{meV} \lesssim m_a \lesssim \SI{100}{meV}$ produces \emph{incoherent} phonons in crystals via its direct interaction with the spin of the nucleons.
Incoherent phonon production dominates when the nuclear spins are randomly oriented, such that no magnetic field is needed.  
The random orientation of the nuclear spins also means that the axion sees a system in which translation invariance is fully broken. 
The axion can thus excite any phonon of frequency $\omega = m_a$, even if the momenta of the axion and the phonon do not match.
The axion can therefore sample all phonon branches over the whole Brioullin zone, leading to a response that is non-resonant and proportional to the atom-projected phonon density of states (see Fig.~\ref{fig:broadbandplot}).  
This allows for a broadband detection concept, requiring only 2 or 3 target materials to cover the 1 meV to 100 meV mass range.

This paper is organised as follows: In~\cref{sec:ratecalc} we calculate the absorption rate. We first recompute the coherent absorption result in~\cite{Mitridate:2020kly, Mitridate:2023izi} and then show how the incoherent process differs. We comment on a similar but subleading source of phonon production from the axion-induced oscillation of the nuclear electric dipole moments (EDM) in~\cref{sec:EDMsection} and review the experimental status and expected backgrounds in~\cref{sec:experiment}. We present our results and comment on future directions in~\cref{sec:results}. The appendices contain further details on some steps in the calculations.

\section{Rate calculation\label{sec:ratecalc}}

We calculate the axion absorption rate in several steps. First, we define the model and isolate the relevant term in the non-relativistic expansion. Next, we match the axion-nucleon operator to the corresponding axion-nucleus operator. Finally, we derive the axion-phonon coupling and compute the coherent and incoherent absorption rates. For the latter case, we average over a fully randomised ensemble of initial state configurations for the nuclear spins in the crystal.

\subsection{Model definition and conventions}
We start from the Lagrangian
\begin{equation}
\mathcal{L}\supset \frac{\alpha_s C_{GG}}{8\pi}\frac{a}{f_a} G^{a\mu\nu}\tilde G_{\mu\nu}^a + \sum_f \frac{C_f}{2f_a}\partial^\mu a \bar f \gamma_\mu\gamma_5 f
\end{equation}
with $f_a$ the axion decay constant and the $C_f$ model-dependent Wilson coefficients associated with the Standard Model fermions $f$. 
We don't consider the axion-photon coupling throughout this work.
At low energies, these operators source axion couplings to the proton and neutron, which can be parametrised by
\begin{equation}\label{eq:protoncoupling}
    \mathcal{L}\supset  g_p\partial^\mu a \bar p \gamma_\mu\gamma_5 p + g_n\partial^\mu a \bar n \gamma_\mu\gamma_5 n.
\end{equation}
The relations between the $g_{n,p}$ and $C_i$ can be found in~\cite{GrillidiCortona:2015jxo}. We will present our results in terms of the $g_{n,p}$ and remain agnostic about the structure of the UV model, unless stated otherwise. The EDM operator induced by the gluon coupling will be discussed in~\cref{sec:EDMsection}.

To match onto the nuclear and ultimately phonon effective theories, we will need to start with the non-relativistic Hamiltonian associated with the axion coupling to the proton/neutron~\eqref{eq:protoncoupling}
\begin{align}
 H_a =   - g_{p,n} (\Bpartial a) \cdot \Bsigma - \frac{g_{p,n}}{m_{p,n}}\, \dot a \Bsigma\cdot\bfD.\label{eq:nonrellimit}
\end{align}
Here, $\Bsigma$ are the Pauli matrices, operating on nucleon spin degrees of freedom, and $\bfD$ is a covariant matrix operating on spatial degrees of freedom.
The procedure to arrive at this result is described in e.g.~\cite{Mitridate:2021ctr,Berlin:2023ubt}. 
The primary effect of the first term is to rotate the nuclear spins in the crystal. For this effect to be observable in a phonon final state, the crystal must have a strong phonon-spin coupling. We leave this case for future work. We instead focus on the second term.
The effect of this operator will be to induce a slight motion of the nucleus in the direction of its spin, effectively creating one or more phonons.

We will only consider systems without large magnetic fields (internal or external), such that we can replace $\bfD\to \Bpartial$ when working in the Coulomb gauge. We further take the positive frequency component of $a$, and so $a=a_0 e^{i m_a t}$ with $a_0=\tfrac{\sqrt{\rho_a}}{\sqrt{2}m_a}$, where we approximate the axion as monochromatic since its width is much narrower than that of a phonon. The operator responsible for axion absorption is then
\begin{align}\label{eq:Hnucleonlevel}
 H_a =  - 2i \frac{g_{p,n}m_a }{m_{p,n}}\, a_0 \bfS_{p,n}\cdot\Bpartial 
\end{align}
with $\bfS_{p,n}$ the spin operator of the proton / neutron.

\subsection{Crystal matrix element}

The Hamiltonian in~\eqref{eq:Hnucleonlevel} leads to an interaction of the axion with the atoms in the crystal,
\begin{align}\label{eq:cmtstart}
  \delta H&= -2 m_a  a_0 g_p \sum_{\bfl j} \frac{\lambda_{\bfl j}}{m_{\bfl j}} \mathbf{k}_{\bfl j}\cdot  \mathbf{J}_{\bfl j},
\end{align}
where $\bfl$ is the position vector of the lattice cell and $j$ is the site in the unit cell at which the atom resides. The $\bfJ_{\bfl,j}$ and $m_{\bfl j}$ represent the spin and mass of the nucleus at position $\bfl,j$. The product $g_p \lambda_{\bfl j}$ is the effective axion-nucleus coupling, which relates to the $g_{n,p}$ through a nuclear matrix element calculation. The overall rate is controlled by $g_p$, while the proportionality constants $\lambda_{\bfl j}$ parametrise the model-dependent $g_p/g_n$ ratio, as well as the nuclear matrix elements. 
This procedure is well known from spin-dependent WIMP scattering and is reviewed in~\cref{app:nuclear}. Here, ${\bf k}_{\bfl j}$ denotes the momentum of the atom at lattice site $\bfl j$. The transition from nuclear momenta to atom momenta is achieved by integrating out the electron degrees of freedom, using the Born-Oppenheimer approximation as discussed in~\cref{app:bo}.

Let $\psi$ and $\psi'$ be two eigenstates of the Hamiltonian describing the spatial and spin degrees of freedom of the atoms in the lattice. We can compute the expectation value
\begin{align}
   \langle \psi' | \mathbf{k}_{\bfl j}\cdot  \mathbf{J}_{\bfl j}|\psi \rangle=& i m_{\bfl j} \langle\psi' | \mathbf{J}_{\bfl j}\cdot [H_0,\mathbf{x}_{\bfl j}]|\psi\rangle
\end{align}
with $H_0 = \sum_{\bfl j} \frac{|\mathbf{k}_{\bfl j}|^2}{2m_{\bfl j}}$ the kinetic part of the Hamiltonian. 
If there are no non-central forces contributing to the potential, in other words no non-negligible magnetic fields, then the potential part of the full Hamiltonian commutes with $\mathbf{x}_{\bfl j}$ and we can replace $H_0$ with the full crystal Hamiltonian $H$ in the commutator
\begin{align}
 \langle \psi' | \mathbf{k}_{\bfl j}\cdot  \mathbf{J}_{\bfl j}|\psi \rangle=&i m_{\bfl j} \langle\psi' |\mathbf{J}_{\bfl j}\cdot [H,\mathbf{x}_{\bfl j}]|\psi\rangle\\
=&i m_{\bfl j} \omega \langle\psi' |\mathbf{J}_{\bfl j}\cdot \mathbf{x}_{\bfl j}|\psi\rangle \label{eq:comrel}
\end{align}
with $\omega\equiv E_{\psi'}-{E_\psi}$, and ${\bf x}_{\bfl j}$ is the position operator of the $\bfl,j$ atom. When commuting the $H$ and $\mathbf{J}_{\bfl j}$ above, we implicitly assumed that wave functions are factorisable as
\begin{equation}\label{eq:factorization}
    |\psi \rangle = \prod_{\bfl j}  |s_{\bfl j} \rangle \otimes | \phi \rangle    
\end{equation}
with $|s_{\bfl j}\rangle$ the spin state of the nucleus indexed by $j,\bfl$ and $|\phi\rangle$ the state describing the spatial degrees of freedom of the atoms in the crystal. This should be a good approximation for non-magnetic materials within the axion frequency range considered here.
This factorisation allows us to write
\begin{align}
\langle \psi' | \mathbf{k}_{\bfl j}\cdot  \mathbf{J}_{\bfl j}|\psi \rangle=&i m_{\bfl j} \omega \langle s'_{\bfl j} |\mathbf{J}_{\bfl j}|s_{\bfl j}\rangle\cdot \langle \phi'|\mathbf{x}_{\bfl j}|\phi\rangle.
\end{align}
Finally, we expand the location of the nucleus around its equilibrium position $\mathbf{x}_{\bfl j}\to \mathbf{x}^0_{\bfl j}+\mathbf{u}_{\bfl j}$, such that 
\begin{equation}\label{eq:matrixelementresult}
\langle \psi' | \mathbf{k}_{\bfl j}\cdot  \mathbf{J}_{\bfl j}|\psi \rangle=i m_{\bfl j} \omega \langle s'_{\bfl j} |\mathbf{J}_{\bfl j}|s_{\bfl j}\rangle\cdot \langle \phi'|\mathbf{u}_{\bfl j}|\phi\rangle.
\end{equation}
for $\phi'\neq\phi$.

To compute the matrix element, we have factorised the initial and final states as product states of the spin and phonon degrees of freedom, as in~\eqref{eq:factorization}, and we now take
\begin{align}
    |i \rangle =& \prod_{\bfl j}  |s^i_{\bfl j} \rangle \otimes | 0 \rangle \\
    |f \rangle =& \prod_{\bfl j}  |s^f_{\bfl j} \rangle \otimes | \nu,\bfk \rangle 
\end{align} 
where $|0\rangle$ is the phonon ground state and $|\nu,\bfk\rangle$ is a single phonon state. The latter is fully specified by the phonon branch, indexed by $\nu$, and the phonon three momentum $\bfk$.
The energy of the phonon $\omega_{\nu \bfk}$ is given by the phonon dispersion relations of the material of interest. 
In general $\omega_{\nu\bfk}$ is a complicated function, which is either measured or calculated using Density Functional Theory (DFT) methods.  
The $\bfu_{\bfl j}$ operator can be quantised as 
\begin{equation}\label{eq:defphonon}
    \mathbf{u}_{\bfl j} = \frac{1}{\sqrt{2 N m_{\bfl j}}} \sum_{\nu\mathbf{k}}\frac{1}{\sqrt{\omega_{\nu\mathbf{k}}}} a_{\nu\mathbf{k}}\Bepsilon_{j\nu\mathbf{k}} e^{i\mathbf{k}\cdot(\mathbf{\bfl}+\mathbf{x}_j^{0})}+\text{h.c.}
\end{equation}
with $N$ the number of crystal cells in the target and $a_{\nu\mathbf{k}}$ ($a_{\nu\mathbf{k}}^\dagger$) the destruction (creation) operator of the phonon with branch $\nu$ and momentum $\bfk$. The $\Bepsilon_{j\nu\mathbf{k}}$ are the phonon polarisation tensors for the atom in site $j$. This leaves us with the following matrix element
\begin{align}
    \label{eq:matrixelementresult}
    \langle f | \delta H |i \rangle =& -2i m_a a_0 g_p\frac{\omega}{\sqrt{2N \omega_{\nu\bfk}}}  \sum_{\bfl j} \frac{\lambda_{\bfl j}}{\sqrt{m_{\bfl j}}} e^{-i \bfk\cdot(\bfl +\bfx^0_j)}\nonumber\\
    &\times \langle s^f_{\bfl j} | \bfJ_{\bfl j} | s^i_{\bfl j} \rangle\cdot \Bepsilon_{j\nu\mathbf{k}} \prod_{\bfl' j' \neq \bfl j} \langle s^f_{\bfl' j'} | s^i_{\bfl' j'}\rangle 
\end{align}
The product over the $\langle s^f_{\bfl' j'} | s^i_{\bfl' j'}\rangle$ encodes that for each term in the sum over $\bfl,j$, the only spin that can be changed between the initial and final states is the spin associated with $\bfl,j$. 
As we will show below, the rate depends in a major way on the nuclear spin expectation values. If all the spins in the crystal are aligned, the terms in the $\bfl,j$ sums will maximally interfere. 
We refer to this process as coherent absorption, in analogy to the terminology used in the context of neutron scattering in crystals.
The coherent absorption was already calculated by Mitridate et.al.~\cite{Mitridate:2023izi}, and we will first rederive their results within our framework. 
We also account for the nuclear form factors, which were neglected in~\cite{Mitridate:2023izi}.

In addition, terms that do not describe interference between two distinct atoms correspond to a qualitatively different process. In~\cref{sec:incoherent} we will show how they describe a non-resonant, broadband absorption process, which we term ``incoherent absorption''.   
If the nuclear spins in the crystal are unpolarised, the interference terms will cancel in the averaging over the initial state configurations, and the incoherent absorption is, therefore, the only surviving process. 

\subsection{Coherent rate calculation}

From the matrix element of~\eqref{eq:matrixelementresult}, we can calculate the absorption rate using Fermi's Golden rule
\begin{align}\label{eq:fermi}
\Gamma (\omega)  =& (2\pi) \sum_{i,f} w_i |\langle i | \delta H | f \rangle|^2 \delta(\omega-\omega_{\nu\bfk}),
\end{align}
where we also average over an ensemble of initial states specified by their spin configurations with weights $w_i$.
Absorption of an axion can lead to coherent phonon production in a perfectly nuclear-spin-polarised crystal, as previously considered in Ref.~\cite{Mitridate:2023izi}. We start with a derivation of this coherent rate to see what restrictions the coherence places on the absorption process. We use the matrix elements derived in the previous section, which we emphasise are only valid in the absence of magnetic fields, for example due to the simplification of~\eqref{eq:comrel}. If keeping the nuclei polarised involves large magnetic fields, our calculation therefore needs to be extended.

We find it useful to split the expression for the rate as follows
\begin{equation}
\Gamma(\omega) = \Gamma_{\text{coh}}(\omega)+\Gamma_{\text{incoh}}(\omega)
\end{equation}
where the $\Gamma_{\text{coh}}(\omega)$ and $\Gamma_{\text{incoh}}(\omega)$ are respectively the off-diagonal and diagonal terms in the double sum over the atoms in the crystal in~\eqref{eq:fermi} (seen by plugging in the expression of~\eqref{eq:matrixelementresult}). Concretely, 
\begin{align}
\Gamma_{\text{incoh}} (\omega)  =& 4\pi m_a^2 \omega a_0^2 g_p^2\frac{1}{N}  \sum_{\bfl j}  \frac{\lambda_{\bfl j}^2}{m_{\bfl j}}  \sum_{i,f} w_i \!\!\!\!\sum_{\alpha,\beta=x,y,z} \!\!\!\!\epsilon_{j\nu\mathbf{k}}^\alpha \epsilon_{j\nu\mathbf{k}}^\beta \nonumber\\
& \times \langle s^i_{\bfl j} |J^\alpha_{\bfl j}|s^f_{\bfl j}\rangle \langle s^f_{\bfl j} |J^\beta_{\bfl j}|s^i_{\bfl j}\rangle 
 \delta(\omega-\omega_{\nu\bfk}), \label{eq:incoherentrate1}
\end{align}
\begin{align}
\Gamma_{\text{coh}} (\omega)  =& 4\pi m_a^2 \omega a_0^2 g_p^2\frac{1}{N}  \sum_{\bfl j}\!\!\! \sum_{\substack{\bfl'\neq \bfl\\\text{ or } j'\neq j}} \!\!\!\frac{\lambda_{\bfl j}\lambda_{\bfl' j'}}{\sqrt{ m_{\bfl j}m_{\bfl' j'}}} \sum_{i,f} w_i  \nonumber\\
& \times \!\!\!\!\sum_{\alpha=x,y,z} \langle s^i_{\bfl j} |J^\alpha_{\bfl j}|s^f_{\bfl j}\rangle \langle s^i_{\bfl j} |s^f_{\bfl j} \rangle \epsilon_{j\nu\mathbf{k}}^\alpha\nonumber\\
& \times \!\!\!\!\sum_{\beta=x,y,z} \langle s^f_{\bfl' j'} |J^\beta_{\bfl' j'}|s^i_{\bfl' j'}\rangle \langle s^i_{\bfl' j'} |s^f_{\bfl' j'} \rangle \epsilon_{j'\nu\mathbf{k}}^\beta
\nonumber\\
&\times   e^{-i \bfk\cdot(\bfl -\bfl')} \delta(\omega-\omega_{\nu\bfk}),
\label{eq:ratebeforefork}
\end{align}
where the sum over the final states $f$ is a sum over the phonon degrees of freedom $\nu,\bfk$ and the spin degrees of freedom. 
In the remainder of this section, we calculate $\Gamma_{\text{coh}} $; we treat $\Gamma_{\text{incoh}}$ in the next section.

Since we neglected the presence of magnetic fields, all spin configurations are degenerate in energy. Therefore, the sum over final state spins is a complete sum over the space of spins, and forms a resolution of the identity. Concretely, the spin matrix elements in~\eqref{eq:ratebeforefork} can be evaluated as 
\begin{align}
    &\sum_{s^f}  \langle s^i_{\bfl j} |J^\alpha_{\bfl j}|s^f_{\bfl j}\rangle \langle s^i_{\bfl' j'} |s^f_{\bfl' j'} \rangle \langle s^f_{\bfl' j'} |J^\beta_{\bfl' j'}|s^i_{\bfl' j'}\rangle  \langle s^f_{\bfl j} |s^i_{\bfl j} \rangle \nonumber\\ &= \langle s^i_{\bfl j} |J^\alpha_{\bfl j}|s^i_{\bfl j}\rangle \langle s^i_{\bfl' j'} |J^\beta_{\bfl' j'}|s^i_{\bfl' j'}\rangle \nonumber\\
    &= J_j J_{j'} \delta^{\alpha z} \delta^{\beta z},
\end{align}
for a crystal whose nuclear spins are fully polarised along the $\hat{z}$ direction, each with the maximal angular-momentum quantum number $J_j$ (i.e.~the spin-states are all independent of $\boldsymbol{\ell}$). For simplicity, we have assumed that for each $j$ there is only a single isotope in the crystal with 100\% abundance. With this assumption, we may also simplify our notations $\lambda_{\boldsymbol{\ell}j} = \lambda_j$, and $m_{\boldsymbol{\ell j}} = m_j$.

 The rate for coherent production now becomes
\begin{align}
\begin{split}
    \Gamma_\mathrm{coh}(\omega) =& 4\pi m_a^2 \omega a_0^2 g_p^2\frac{1}{N} \sum_{\nu,\mathbf{k}} \sum_{ j j'} \frac{\lambda_{j}\lambda_{ j'}}{\sqrt{ m_{ j}m_{j'}}} J_j J_{j'} \\
    &\times \epsilon^z_{j \nu \mathbf{k}} \epsilon^z_{j' \nu \mathbf{k}} \sum_{\boldsymbol{\ell} \boldsymbol{\ell '}} e^{-i\mathbf{k} \cdot (\bfl - \bfl')} \, \delta(\omega - \omega_{\nu \mathbf{k}}) \\
    =& 4\pi m_a^2 \omega a_0^2 g_p^2 N \sum_{\nu, \mathbf{k}} \Big| \sum_j \frac{\lambda_j J_j \epsilon^z_{j \nu \mathbf{k}}}{\sqrt{m_j}} \Big|^2 \delta_{\mathbf{k},\mathbf{0}} \delta(\omega- \omega_{\nu \mathbf{k}}),
    \label{eqn:coherentRate}
\end{split}
\end{align}
where in the second equality we've used the identity $\sum_{\bfl} e^{-i(\mathbf{k} - \mathbf{p}) \cdot \bfl } = N \delta_{\mathbf{k},\mathbf{p}}$ twice. We see the coherent result effectively constrains $|\mathbf{k}|=|\mathbf{p}_a|\ll$ keV\footnote{For simplicity, we dropped the spatial dependence of the axion field, but had we kept the axion momentum, the $\delta_{{\bf k},0}$ would have instead been $\delta_{{\bf k},{\bf p}_a}$.}, imposing that the phonon momentum must match the axion momentum. This is a direct result of the alignment of all the spins, which is a configuration that preserves the long-distance translation symmetry of the crystal lattice. As a result, $|\mathbf{k}|$ is very small compared to the size of the Brillouin zone.

For the acoustic modes, the dispersion relations then imply that there is no allowed phase space for the absorption process.
The dispersion relations of the optical phonons on the other hand are to good approximation independent of $\bfk$ for $\bfk\ll$ keV. They can therefore be excited through coherent absorption, where one can take $\bfk\approx 0$ as in~\eqref{eqn:coherentRate}.
This rate is divergent on-resonance $\omega = \omega_{\nu,\mathbf{k}}$ since we have taken the excited phonon states to be absolutely stable and the axion to be perfectly monochromatic. In reality, the phonons have some width $\gamma_{\nu,\mathbf{k}}$ which is larger than the axion width, and we approximate the regulated result as a simple Breit-Wigner distribution
\begin{align}2\pi \delta(\omega - \omega_{\nu,\mathbf{k}}) \to \frac{ \gamma_{\nu,\mathbf{k}}}{(\omega - \omega_{\nu,\mathbf{k}})^2 + \gamma_{\nu,\mathbf{k}}^2},
\end{align}
although the true off-resonant behavior will be more complicated, and depends on the microscopic phonon dynamics~\cite{Campbell-Deem:2019hdx}. As we only wish to compare the resonant rate to the incoherent case, using a more accurate form is beyond this work. Regardless, we see that~\eqref{eqn:coherentRate} for the coherent rate has resonant sensitivity to an axion whose mass matches the frequency $\omega_{\nu,\mathbf{0}}$ of a zero-momentum phonon.

Since the rate is an extrinsic quantity depending on the size of the detector, we define an intrinsic rate per unit mass as
\begin{align}\label{eq:Rdefinition}
    R(m_a) \equiv \frac{\Gamma_{\text{coh}}(m_a)}{N \times \sum_j m_j}.
\end{align}
The quantity $R$ describes the number of events per unit (mass$\times$time), and so captures the intrinsic absorption rate for a given material. For the particular case of diatomic crystals, such as \ch{H2} or  GaAs, we find in~\cref{app:coherent} a simple analytic formula for the coherent excitation rate of an optical phonon with frequency $\omega_\mathrm{O}$
\begin{align}
    R_\mathrm{O}(m_a) = \frac{ \rho_a g_p^2 (\lambda_1 J_1 - \lambda_2 J_2)^2}{3(m_1 + m_2)^2}   \frac{\omega_{\mathrm{O}}\,\gamma_{\mathrm{O}}}{(\omega_{\mathrm{O}} - m_a)^2 + \gamma_{\mathrm{O}}^2},
    \label{eqn:coherentRateLO}
\end{align}
where we are exciting an optical phonon whose polarisation vector points along the direction of the spin polarisation. We see the narrowband sensitivity to axions with $m_a \approx \omega_\mathrm{O}$, provided $\lambda_1 J_1 \neq \lambda_2 J_2$.

This rate differs from the result previously found in the literature~\cite{Mitridate:2023izi}, as we include the $\lambda_i$ factors that arise when matching from nucleons to nuclei. However, setting $\lambda_i = 1$, and assuming that every nucleon spin in a nucleus is aligned, we recover the rate as previously forecast~\cite{Mitridate:2023izi}. In reality, such a state is a highly excited nuclear state with energy $\gg \SI{}{MeV}$, making it challenging to create in a macroscopic, stable configuration. A more achievable scenario is to have a spin-polarised crystal with each nucleus in its nuclear ground state. For GaAs\footnote{A natural GaAs crystal contains different isotopes of gallium -- as such, we average over the isotopic abundance in calculating $\lambda$, as described in~\cref{app:isotopes}.}, for instance, the protons have $\lambda_\mathrm{Ga} J_\mathrm{Ga} = 0.15$ and $\lambda_\mathrm{As} J_\mathrm{As} = -0.01$, and the expected rates are reduced by two orders of magnitude for the proton coupling. The rate is further suppressed for the neutron coupling, since both Ga and As have an even number of neutrons (similarly for the $^{27}$Al and $^{16}$O in sapphire). Finally, we note that Eq.~\eqref{eqn:coherentRateLO} also applies to the two transverse optical phonon modes, which may also be excited in addition to the longitudinal optical mode.

\subsection{Incoherent rate calculation}
\label{sec:incoherent}
Next, we calculate the incoherent contribution in~\eqref{eq:incoherentrate1}. 
We will now assume a crystal where all nuclear spins are oriented randomly, which is the more straightforward case from an experimental point of view. The calculation for the polarised case is largely analogous and gives a comparable $\Gamma_{\rm incoh}$.
Since it is impossible to know the exact initial state configuration of all the spins in the crystal, we instead calculate the rate as averaged over an ensemble of all possible initial spin configurations. We use the weights $w_i$ to implement this ensemble average. The difference between the ensemble-averaged rate and the rate for a generic initial configuration should go to zero as $N\to\infty$.

The ensemble-averaged coherent rate in~\eqref{eq:ratebeforefork} will vanish for an unpolarised crystal
because the average over all possible $i$ will ensure that the $\langle s^i_{\bfl j} |J^\alpha_{\bfl j}|s^f_{\bfl j}\rangle \langle s^f_{\bfl' j'} |J^\beta_{\bfl' j'}|s^i_{\bfl' j'}\rangle$ correlators cancel for $\bfl\neq\bfl'$ or $j\neq j'$. 
We are therefore just left with the incoherent absorption rate in~\eqref{eq:incoherentrate1}, which we repeat here for the convenience of the reader
\begin{align}
\Gamma (\omega)  =& 4\pi m_a^2 \omega a_0^2 g_p^2\frac{1}{N}  \sum_{\bfl j}  \frac{\lambda_{\bfl j}^2}{m_{\bfl j}} \sum_{\alpha,\beta=x,y,z} \sum_{i,f} w_i  \nonumber\\
& \times \langle s^i_{\bfl j} |J^\alpha_{\bfl j}|s^f_{\bfl j}\rangle \langle s^f_{\bfl j} |J^\beta_{\bfl j}|s^i_{\bfl j}\rangle 
\epsilon_{j\nu\mathbf{k}}^\alpha \epsilon_{j\nu\mathbf{k}}^\beta \delta(\omega-\omega_{\nu\bfk}) 
\label{eq:incoherentstart}
\end{align}
In contrast to the coherent case in~\eqref{eqn:coherentRate}, the $e^{-i \bfk\cdot(\bfl -\bfl')}$ phase canceled in~\eqref{eq:incoherentstart}, removing the $\bfk=0$ constraint on the phase space. Physically, this happens because the random orientations of the nuclear spins imply that the axion sees a target in which the translation symmetry of the phonon effective theory is fully broken. The phonon momentum is therefore not constrained by the axion momentum in this absorption process, and we can sample the full Brillouin zone through the sum over $\bfk$.

We again assume that spin configurations have degenerate energies.\footnote{As we saw in~\eqref{eq:matrixelementresult}, the axion interaction can only flip a single spin to leading order in $g_p$, and so the only final spin states that contribute to the rate have at most one spin flip -- that associated with spin $\bfl, j$. This justifies approximating the spin states as degenerate.} This means that the sum over the $|s^f_{\bfl j}\rangle \langle s^f_{\bfl j} |$ in~\eqref{eq:incoherentstart} is a sum over a complete set of eigenstates, and we can replace it with the identity
\begin{align}
\Gamma(\omega) =& 4\pi m_a^2 \omega a_0^2 g_p^2 \frac{1}{N} \sum_{\bfl j}  \frac{g_{\bfl j}^2}{m_{\bfl j}} \sum_{\alpha,\beta=x,y,z}\sum_{i} w_i  \nonumber\\
&   \langle s^i_{\bfl j} |J^\alpha_{\bfl j} J^\beta_{\bfl j}|s^i_{\bfl j}\rangle  \sum_{\nu,\bfk}  \epsilon_{j\nu\mathbf{k}}^\alpha \epsilon_{j\nu\mathbf{k}}^\beta  \delta(\omega-\omega_{\nu\bfk}). \label{eq:incoherentsum}
\end{align}

When assuming a randomised ensemble, one can moreover show that
\begin{equation}
 \frac{1}{N}\sum_{\bfl}\sum_{i} w_i  \frac{\lambda_{\bfl j}^2}{m_{\bfl j}} \langle s^i_{\bfl j} |J^\alpha_{\bfl j} J^\beta_{\bfl j}|s^i_{\bfl j}\rangle = \frac{1}{3} \xi_j \delta^{\alpha\beta}
 \label{eqn:ensembleAverage}
\end{equation}
with
\begin{equation}
    \xi_j\equiv \frac{1}{N}\sum_{\bfl}\frac{\lambda_{\bfl j}^2}{m_{\bfl j}} J_{\bfl j}(J_{\bfl j}+1).
    \label{eqn:xij}
\end{equation} 
We show the relation in~\eqref{eqn:ensembleAverage} in~\cref{app:angularaverage}; \cref{tab:xi} contains its numerical values for the nuclei we consider. We can further define the partial density of states  $D_j(\omega)$ as\footnote{We assume that the presence of different isotopes does not significantly alter the phonon spectrum. This assumption is valid if a single isotope has a much larger abundance than the others, or if the mass difference between the isotopes is small relative to their average mass.} 
\begin{equation}
  D_j(\omega)\equiv  \frac{1}{3N } \sum_{\nu\mathbf{k}}|\Bepsilon_{j\nu\mathbf{k}}|^2\delta(\omega-\omega_{\nu\mathbf{k}}).
\end{equation}
For most materials, the $D_j(\omega)$ has been calculated with DFT methods, and/or measured with neutron scattering.
The rate is then
\begin{align}\label{eq:absoluterate}
\Gamma(\omega) =& 4\pi m_a^2 \omega a_0^2  g_p^2 N \sum_{j}  \xi_j   D_j(\omega). 
\end{align}
We obtain the density of states from prior DFT calculations and/or measurements. We consider $\text{H}_2\text{O}$~\cite{icedos} and $\text{D}_2\text{O}$~\cite{icedos}\footnote{Ref.~\cite{icedos} only shows the acoustic, optical and libration (rotation) modes of the two water (or heavy water) molecules in the primitive cell, which corresponds to only 12 of the 18 vibrational degrees of freedom. We therefore multiply their density of states by $2/3$ to remain consistent with our normalisation. The remaining modes live at higher frequencies and are not included in our analysis.}, solid $\text{H}_2$~\cite{10.1063/1.1649312}, solid $\text{D}_2$~\cite{PhysRevB.80.064301}, $\text{Li}_2\text{O}$~\cite{2004Prama..63..409G}, sapphire ($\text{Al}_2\text{O}_3$)~\cite{Knapen:2021bwg} and Be~\cite{Bary}.

\begin{table}
\begin{tabular}{ll  } 

  &  $\xi_j\; [\SI{}{GeV}^{-1}]$ \\
 \hline
 $^{1}$H & 0.75  \\
 $^2$D & $0.21\times(1+g_n/g_p)^2$ \\
 $^7$Li & \num{3.3e-2}  \\
 $^9$Be & \num{1.8e-2}$\times(g_n/g_p)^2$\\
 $^{27}$Al$\quad$ & \num{3.9e-3}$+$ \num{6.4e-5}$\times(g_n/g_p)^2$ \\
 $^{69}$Ga & \num{2.7e-4} \\
 $^{71}$Ga & \num{1.2e-3} \\
 $^{75}$As & \num{2.5e-6} 
 \end{tabular}
 \caption{A list of the isotopes with non-zero spin contained in the materials we consider, along with their $\xi_j$ parameter, defined in~\eqref{eqn:xij}. See~\cref{app:angularaverage} for details.  }
 \label{tab:xi}
 \end{table}

Lastly, all that remains is to plug in the axion amplitude $a_0^2=\rho_a/2m_a^2$ and to substitute $\omega \to m_a$. Converting this result into units of events per kg-year exposure, using~\eqref{eq:Rdefinition}, yields
\begin{equation}\boxed{
   R=2\pi \frac{\rho_a m_a}{\overline{\sum_j m_j}} g_p^2   \sum_{j} \xi_j D_j(\omega)}\label{eq:finalanswer}
\end{equation} 
To correctly interpret this result, it is important to recall that the sums over the $j$ are sums over different positions within the unit cell. 
In some materials, some elements occur multiple times in the unit cell, and should be counted as such. 
For example, the unit cell of sapphire $\mathrm{Al}_2\mathrm{O}_3$ has 4 aluminum and 6 oxygen atoms. 
The most abundant oxygen isotope, $^{16}$O, has spin 0 and therefore does not contribute to the rate, but it must still be accounted for in the $\overline{\sum_j m_j}$.
Likewise, the Al atoms must be accounted for with multiplicity 4 in the $\sum_{j} \xi_j D_j(\omega)$ summation. Therefore, in most scenarios, the multiplicity should cancel out between these two sums.

From~\eqref{eq:finalanswer} we see that the rate depends directly on the mass of the nuclei through the coupling $\xi_j$ and indirectly through the density of states. 
Therefore, for a fixed detector mass, low mass nuclei are preferred, as is evident from the denominators in~\eqref{eq:finalanswer} and~\eqref{eqn:xij}. 
Finally, the rate is proportional to the density of states $D_j(\omega)$, as we can sample the whole Brillouin zone to match $\omega$ to the phonon frequency $\omega_{\nu\bfk}$.
Naturally, the rate is still largest near the frequencies of the optical phonons, as those tend to have flatter dispersion bands than the acoustic branches. 
Unlike in the coherent case, the incoherent rate is not dependent on highly suppressed multiphonon processes if $m_a\neq \omega_{\nu \bfk}$ for $\bfk=0$.
In this sense, the incoherent absorption is a ``broadband'' process, though within the frequency range in which the target material supports phonon modes.

We also comment that even in the previous case where all the initial state spins are polarised, there still exists a contribution to the rate from the $\bfl = \bfl'$ terms in the sum in~\eqref{eq:ratebeforefork}. This contribution,  from the $xy$ components of  ${\bf J}$, flips individual spins, and excites modes across the Brillouin zone. In this case, the breaking of translation invariance in the phonon effective theory arises from the random reorientation of the \textit{final state} spins. The corresponding excitation rate is broadband, and numerically similar to the rate~\eqref{eq:finalanswer} in an unpolarised crystal.

\section{Induced time-dependent EDM \label{sec:EDMsection}}

Axions can also induce a time-dependent electric dipole for the nuclei through the $\frac{a}{f_a}\tilde G G$ operator.
There are two possibilities for how this can be generated, either through the intrinsic EDMs of the nucleons, which are generally dominated by the spin-unpaired valence nucleon, or by parity (P) and time reversal (T) violating nuclear potentials, which can induce an asymmetric distribution of charge within the nucleus. In the former case, the QCD axion induces the following EDMs for individual neutrons and protons (see e.g.~\cite{Yamanaka:2017mef, Pospelov:2005pr}):
\begin{align}
    d_p(t) &= (2.1 \pm 1.2) \times 10^{-16}\times \frac{a(t)}{f_a} e\,\rm{cm}\\
    d_n(t) &= -(2.7 \pm 1.2) \times 10^{-16} \times \frac{a(t)}{f_a} e\,\rm{cm}
\end{align}
Like in the case of the $g_{ann}$ operator described in~\cref{sec:ratecalc}, the nuclear EDM can be found through the Wigner-Eckhart theorem, as described in~\cref{app:nuclear}:
\begin{equation}
    H_{EDM} = \mathbf{E}_N \cdot \left[d_p \sum_{i=1}^{Z}\frac{\mathbf{S}_{p_i}}{S_{p_i}} + d_n\sum_{i=1}^{A-Z}\frac{\mathbf{S}_{n_i}}{S_{n_i}} \right]\label{eq:nucleonedm}
\end{equation}
We want to match \eqref{eq:nucleonedm} onto the nuclear EDM operator:
\begin{equation}
    H_{EDM}=d_{(1)} \,\mathbf{E}_N\cdot \, \frac{\mathbf{J}}{J}.
\end{equation}
This can be achieved using the Wigner-Eckhart theorem, giving the estimate:
\begin{equation}
    d_{(1)} =  2 J \,\left(\lambda_p d_p + \lambda_n d_n\right)
\end{equation}
where $\lambda_{p,n}$ are defined in equations \eqref{eq:deflambdap} and \eqref{eq:deflambdan}.

The contribution due to PT violating forces is more complicated. Here the dominant effect is due to one-pion exchange between nucleons~\cite{Khriplovich:1997ga, Flambaum:1984fb, deVries:2020iea}.  
An analytic estimate applicable to heavy nuclei is reviewed in~\cref{app:PTviolatingpot}. For the light nuclei featured in this paper, however, this approximation is highly inaccurate, and we instead refer to the results of the more detailed calculations found in \cite{Engel:2013lsa, Yamanaka:2019vec} and references therein. Both contributions of the intrinsic EDMs and pion-exchange forces are taken into account here, and the light nuclei $^2$D, $^3$He, $^7$Li, and $^9$Be are all found to have EDMs of order
\begin{equation}\label{eq:EDMlight}
    d = d_{(1)}+d_{(2)} \sim 10^{-16}\, \frac{a(t)}{f_a}\,\rm{e\, cm},
\end{equation}
where $d_{(2)}$ signifies the contribution from PT violating forces. From here on, we will also define the proportionality constant $d_\theta$ by $d(t) \equiv d_\theta \frac{a(t)}{f_a}$. For heavy nuclei, the estimates of~\cref{app:PTviolatingpot} suggest the contribution from the PT violating forces could be an order of magnitude larger than~\eqref{eq:EDMlight}.
However, as we will see below, other suppression factors for heavy nuclei more than offset this enhancement, and we therefore do not consider them any further.

A time-dependent dipole moment can induce dipole transitions in the electronic shells of the atom~\cite{Flambaum:2019rvu}.
We show here that at lower frequencies, it also sources incoherent phonons. 
The EDM operator for the nuclei in the crystal is 
\begin{equation}
\delta H_{\rm EDM}=\sum_{{\bfl}j}d_{{\bfl}j}(t){\bf E}_{{\bfl}j }\cdot {\bf \hat{J}}_{{\bfl}j},
\label{eq:deltaH_EDM}
\end{equation}
with ${\bf E}_{{\bfl}j}$ as the electric field at the site of the nucleus, ${\bf \hat{J}}_{{\bfl}j}$ as the unit vector of the spin of the nucleus, and $d_{{\bfl}j}$ is the contribution to the EDM of the nucleus, induced by the axion.

We will show that~\eqref{eq:deltaH_EDM} induces the same transition matrix elements as direct coupling to the nuclear spin in~\eqref{eq:cmtstart}, albeit with a different coefficient. This can be seen as follows: In the absence of sizable magnetic fields, we note that the electric field experienced by the ion in location $\bfl j$ can be written as
\begin{equation}
{\bf E}_{{\bfl}j}=i\frac{1}{eZ_{{\bfl} j}}[H,{\bf k}_{{\bfl} j}].\label{eq:Ecomrel}
\end{equation}
with $Z_{\bfl j}$ the charge of the ion and $H$ the Hamiltonian of crystal, unperturbed by the axion.
Using~\eqref{eq:Ecomrel}, the matrix element is therefore
\begin{align}
    \bra{i}\delta H_{\rm EDM}\ket{f} =& i  \sum_{{\bfl}j} \frac{d_{{\bfl}j}(t)}{e Z_{{\bfl}j} }\,\langle f |[H,{\bf k}_{{\bfl} j}]\cdot {\bf \hat{J}}_{{\bfl}j}| i \rangle\\
    =& i \omega  \sum_{{\bfl}j}   \frac{d_{{\bfl}j}(t)}{e Z_{{\bfl}j}}\,\langle f | \mathbf{k}_{\bfl j}\cdot \mathbf{\hat{J}}_{\bfl j}| i \rangle.\label{eq:EDMprocessed}
\end{align}
We notice that the operator in~\eqref{eq:EDMprocessed} is the same as that for direct coupling to the nuclear spin in~\eqref{eq:cmtstart}.

If we compare the matrix elements term by term, their ratio scales as 
\begin{equation}
\frac{\bra{i}\delta H_{\rm EDM}\ket{f} }{\bra{i}\delta H\ket{f} } \sim \frac{i}{2 e f_a g_p}\times \frac{d_{\theta{\bfl}j} m_{{\bfl}j}}{Z_{{\bfl}j}J_{{\bfl}j} \lambda_{\bfl j}}.
\end{equation}
There is some model dependence on how $g_{n,p}$ relate to $f_a$. One however usually expects their product to be at most of order unity. For example, for the KSVZ axion one finds~\cite{GrillidiCortona:2015jxo}
\begin{equation}
    g_p f_a \approx - 0.24(2) \quad\mathrm{and}\quad g_n f_a \approx - 0.01(2).
\end{equation}
For hydrogen nuclei in the KSVZ model, the magnitude of the ratio of matrix elements above is around $\sim 2\times 10^{-2}$. Given that the rate depends on the matrix element squared, this suggests that the EDM coupling provides weaker sensitivity to the QCD axion in many standard UV completions. For this reason, we have focused on the model-dependent axion-nucleon couplings in this work. For completeness, we obtain the rate of incoherent phonon production produced by the EDM coupling by rescaling the rate given in~\eqref{eq:finalanswer}:
\begin{align}
   R&=\frac{ \pi \rho_a m_a}{2 \overline{\sum_j m_j}e^2 f_a^2}  \sum_{j} \kappa_j D_j(\omega)\label{eq:secondfinalanswer}
   \\
    \kappa_j &= \frac{1}{N} \sum_{\bfl} \frac{d_{\theta \bfl j}^2 m_{\bfl j}}{Z_{\bfl j}^2}\frac{J_{\bfl j}+1}{J_{\bfl j}}.
\end{align}

\section{Backgrounds and experimental status\label{sec:experiment}}

Detection of single phonon excitations below $\sim$ eV energies is a newly emerging frontier in direct detection experiments. 
Currently, the two most well-developed technologies for this purpose are TESs~\cite{Fink:2020noh} and KIDs~\cite{golwala2008wimp, Moore:2012au, Cardani:2021iff, Cruciani:2022mbb, Temples:2024ntv}. We refer to \cite{Essig:2022dfa} for a recent review of other proposed detection schemes.

In this work, we have envisioned an experimental setup similar to that already in development for the TES-based SPICE experiment~\cite{Knapen:2017ekk,Hochberg:2015fth}, which is part of the TESSERACT collaboration. 
The SPICE experiment consists of a cryogenically-cooled target crystal attached to a superconducting aluminium fins, which are in contact with an array of TES. Athermal phonons excited within the target crystal can break Cooper pairs within the aluminium on a timescale faster than the excitation thermalises within the crystal. 
The subsequent change in resistance of the aluminium is detected by the TES.  
The threshold energy for a phonon to break a Cooper pair in aluminium is $7.2$ meV, setting a theoretical lower limit for the axion mass that can be detected. 
The SPICE experiment will use $\text{Al}_2\text{O}_3$ and GaAs crystals; however, in this work, we have also considered materials with lower Z nuclei to enhance the axion absorption rate.

Understanding the backgrounds of solid-state detectors at the energy scales relevant to this work is an active area of research. The growth of extremely pure crystals with minimal radioactive impurities and substantial detector shielding has already been demonstrated in higher energy-threshold experimental searches for DM scattering and ionisation, such as SuperCDMS~\cite{Zatschler:2024coe} and EDELWEISS~\cite{Edelweiss:2022bzh}. There are several possible major backgrounds in the meV energy range. One possibility is low-energy phonons generated by Rayleigh scattering of high-energy photons~\cite{Berghaus:2021wrp}. Another one is the generation of phonons through the absorption of infrared photons. These photons may be created through Cherenkov radiation or radiation from the recombination of electron-hole pairs produced by high-energy particles, such as gamma rays~\cite{Du:2020ldo}. The rates associated with these processes are highly material-dependent. The background from Rayleigh scattering was estimated for a number of materials in Ref.~\cite{Berghaus:2021wrp}. For GaAs, the peak rate in a 2 meV-wide bin in the energy range of interest was $\sim 10$ events per kg year  and for Al$_2$O$_3$ $\sim0.1$ events per kg year. Since this process involves electron scattering, it is likely significantly reduced for the low-Z nuclei most relevant for this work. These backgrounds could be mitigated by using an active veto with timing information that looks for a high-energy event correlated with the low-energy one. The veto could also be calibrated to mitigate backgrounds from phonons caused by relaxing defects that were initially generated by high energy events, by finding the associated relaxation time constant and energy scale. A veto system that uses simultaneous photon and phonon detection using a TES was already demonstrated in Ref.~\cite{Romani:2024rfh}.

At this time, low threshold calorimetric DM-detectors see a large, unexplained excess of low energy events~\cite{Fuss:2022fxe}. 
This low energy excess (LEE) is at least in part due to so-called “quasi-particle poisoning”, i.e.~the breaking of Cooper pairs in the superconducting films. 
Proposed sources of this poisoning include stray radiation, and stress-induced phonons, the latter of which was recently shown to be reduced by suspending the detectors rather than mounting them with glue~\cite{Anthony-Petersen:2022ujw}. 
This has now been further validated with an experiment that was read out with two TES channels~\cite{Romanitalk,Anthony-Petersen:2024vdh}.
This experiment has demonstrated that the LEE has multiple sources: Some events only show up in a single channel, indicating that they are likely related to the relaxation of stress in the corresponding film. 
Other events deposit roughly equal energy in both channels, indicating that they originate from the bulk of the substrate.
Multichannel discrimination alone will therefore not be sufficient to reduce the backgrounds to the level typically associated with DM applications, however, it provides a powerful handle to further understand the LEE, and thus advance the R\&D into this difficult problem.

In summary, major progress has been made towards understanding and reducing the backgrounds in low threshold calorimeters over recent years. 
The most serious challenge at the moment is to mitigate or discriminate against phonon bursts, thought to be due to the relaxation of stress in the detector. 
As we shall see in the next section, achieving access to the allowed parameter space for the QCD axion requires reducing background rates below a few events per kg-year exposure. Increasing the fiducial mass of existing detectors by a few orders of magnitude is already a part of the TESSERACT road map~\cite{TESSERACT_snowmass}. 
Although reaching the necessary high exposure and low background rate still presents major challenges, steady progress continues in this direction, bringing us closer to the conditions needed to probe the QCD line.

\section{Results and discussion\label{sec:results}}

\begin{figure*}
    \centering
    \includegraphics[width=0.5\linewidth]{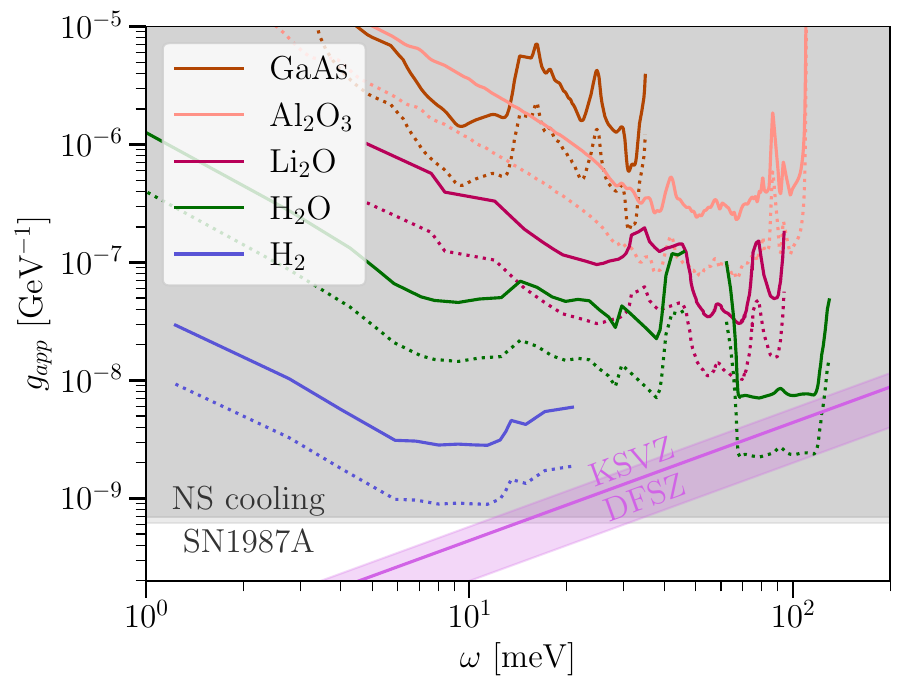}\hfill
    \includegraphics[width=0.5\linewidth]{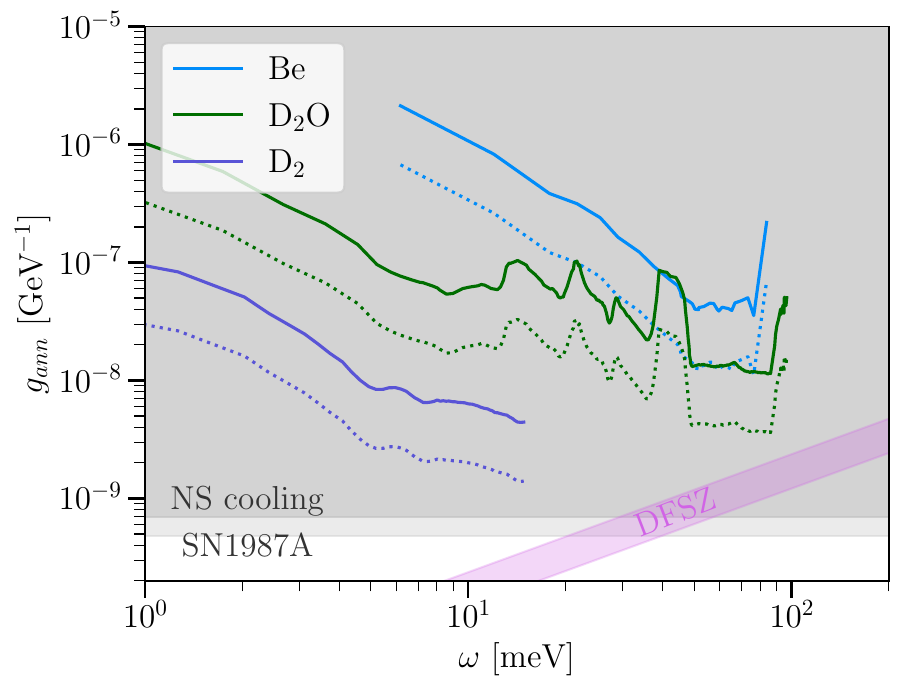}
    \caption{Rate contours for the axion-proton (left) and axion-neutron coupling (right), corresponding to a signal rate of 3 events per kg yr (3 events per 10 kg yrs), as denoted by the solid (dashed) lines. 
   These lines only correspond to projected 95\%C.L. bounds under the assumption of negligible backgrounds, which has not yet been demonstrated experimentally.
    The rate is largest for materials consisting of light nuclei, which also have an odd number of protons (left panel) or an odd number of neutrons (right panel). Bounds coming from the cooling of neutron stars are labeled `NS cooling'~\cite{Buschmann:2021juv}. The bounds from the duration of the neutrino burst of SN 1987A are shown as `SN 1987A', where we conservatively neglect the effect of pions in the neutron star core \cite{Carenza:2019pxu,Lella:2023bfb}.}
    \label{fig:neutron}
\end{figure*}

The solid (dashed) lines in the left-hand panel of Fig.~\ref{fig:neutron}  indicate the axion-proton coupling needed for a rate of 3 signal events per 1 kg-year (10 kg-year) exposure, assuming a local axion density of $\rho_a = \SI{0.4}{GeV.cm^{-3}}$. As anticipated in~\cref{sec:incoherent}, low-mass targets offer the best performance. For comparison, we also show the heavier targets ${\rm Al}_2{\rm O}_3$ and GaAs, as these are some of the default materials that will be used by the SPICE experiment. Ice, in comparison, performs better and has been identified as a promising target for models where dark matter produces phonons through scattering \cite{Taufertshofer:2023rgq}. Solid H$_2$ performs best overall, since hydrogen has the highest spin-to-mass ratio of any element. The right-hand panel of Fig.~\ref{fig:neutron} is analogous to the left-panel, but for the axion-neutron coupling. 
In this case, we see that solid $\text{D}_2$ and $\text{D}_2 \text{O}$ have the largest rates, since these are the lightest targets with an unpaired neutron. 

As shown in~\cref{sec:EDMsection}, an axion would also induce phonon production through the EDM it generates for the nuclei in the crystal. However, in generic QCD axion models, this effect is consistently smaller than the direct coupling to nuclear spin and was therefore not shown in Fig.~\ref{fig:neutron}.

Even in an ideal background-free experiment, a large exposure is needed to achieve sensitivity in regions unconstrained by astrophysical observations or to explore the QCD axion prediction. However, TESSERACT has been funded and approved for construction, and no other competitive terrestrial bounds exist in this mass range. In light of this, it would be worthwhile for the collaboration to pursue a search for this signal. We emphasise that such a search has broadband sensitivity, covering a range of frequencies wherever the phonon density of states has support. This means that no experimental parameters need to be modified during data collection to scan across a wide frequency range.

In the case of axions, it appears challenging to improve upon existing astrophysical constraints. This difficulty arises because the rate in~\eqref{eq:finalanswer} is suppressed by a factor $m_a/m_N$, typical for processes involving axion coupling to the spin of Standard Model fermions. However, it may be that the axion-wind term in~\eqref{eq:protoncoupling} has a more favorable scaling. The effect of this coupling is to rotate nuclear spins, without exciting phonon degrees of freedom at leading order in the axion momentum expansion.
It may be possible to identify a suitable material with a strong phonon-spin coupling, in which, such a spin rotation may be detectable with a low threshold calorimeter. Thus far, we have only considered systems where this coupling was explicitly assumed to be weak; it would be interesting to expand these computations by relaxing that assumption.

More generally, we learned that disorder in the target results in a broadband rather than narrowband absorption spectrum. This is because the disorder fully breaks the translation invariance in the phonon effective theory, such that the axion momentum does not need to match the momentum of the phonon that is created. This is in contrast to coherent absorption, which only occurs in ordered crystals, and for which the phonon momentum must match the axion momentum. 
Disorder is most intuitively present in crystals whose initial state configuration does not respect the crystal's translation symmetries, e.g.~through the random orientation of its nuclear spins. 
It suffices however that either the initial or the final state has this property. For example, a fully polarised crystal can still have an incoherent absorption rate, provided that the interaction flips one (or more) nuclear spins.
In this paper, we focused on nuclear-spin-induced disorder, but DM which couples to systems with other forms of disorder could produce similar results. We leave this exploration to future work.

\section{Acknowledgements}
We thank  Asimina Arvanitaki, Daniel Carney, Stefania Gori, Sin\'ead Griffin, Tongyan Lin, Pankash Munbodh, Nick Rodd, Bethany Suter, Tanner Trickle, Kevin Zhang and Kevin Zhou for useful discussions. We especially thank Pankash Munbodh for detailed discussions about the nuclear form factors and Tanner Trickle for comments on the manuscript. We also thank Daniel Carney for collaboration during the early stages of the project.
The work of IB and SK was performed in part at Aspen Center for Physics, which is supported by National Science Foundation grant PHY-2210452 and the Simons Foundation (1161654, Troyer). AM is supported by grant NSF PHY-2309135 to the Kavli Institute for Theoretical Physics (KITP) and grant 7392 from the Moore Foundation. Research at the Perimeter Institute is supported in part by the Government of Canada through the Department of Innovation, Science and Economic Development Canada and by the Province of Ontario through the Ministry of Colleges and Universities. 
This work is supported in part by the Office of High Energy Physics of the U.S.\ Department of Energy under contract DE-AC02-05CH11231. 

\bibliography{main.bib}

\appendix 

\section{Nuclear form factors\label{app:nuclear}}
In this appendix, we review how the proportionality constants $\lambda_{\ell j}$ are being computed. For completeness, we also briefly discuss the effect of P \& T violating nuclear potentials for large nuclei, which is in principle relevant for the EDM interaction in~\cref{sec:EDMsection}.

\subsection{Nuclear spin form factors}
To go from the axion-nucleon coupling in~\eqref{eq:Hnucleonlevel} to the axion-nucleus effective interaction in~\eqref{eq:cmtstart}, we must account for the nuclear form factors. Since the axion wavelength is much larger than the size of the nucleus, we can add the contributions of the nucleons coherently
\begin{equation}
    H= -2i\frac{m_a }{m_p} a_0\sum^A_{i=1} g_i \bfS_i \cdot \Bpartial_i 
\end{equation}
where we took $m_p\approx m_n$ and defined $A$ to be the mass number of the nucleus. We can perform a coordinate transformation to write this as
\begin{equation}\label{eq:Hamnucleonsum}
    H= -2i\frac{m_a }{m_N} a_0\left[g_p \sum^Z_{i=1}  \bfS_{p_i}+ g_n \sum^{A-Z}_{i=1}  \bfS_{n_i} \right] \cdot \Bpartial_{N} + \cdots
\end{equation}
where we separated the sums over the protons and neutrons. $m_N=A m_p$ is the mass of the nucleus and $\Bpartial_{N}$ is the spatial derivative with respect to the center-of-mass coordinate of the nucleus. The $\cdots$ represent terms involving the internal degrees of freedom of the nucleus and will be dropped in the remainder of the calculation.

Next, we must match~\eqref{eq:Hamnucleonsum} to an effective Hamiltonian involving the spin of the nucleus ($\bfJ$). We will only consider nuclei with non-zero spin, and assume that the contributions from zero spin nuclei, e.g.~$^{16}O$, are negligible. From the Wigner-Eckhart theorem, we know that the matrix elements of the $\sum^Z_i  \bfS_{p_i}$ and $\sum^{A-Z}_i  \bfS_{n_i}$ must be proportional to the matrix element of $\bfJ$, with a constant which is independent on the magnetic quantum number of the spin state~\cite{Engel:1989ix,Engel:1992bf}. 
This means that we can define an effective Hamiltonian of the form
\begin{equation}\label{eq:Hamnuclear}
    H= - 2\frac{m_a }{m_N} a_0 g_N \bfJ \cdot \bfk_N
\end{equation}
where we also switched to momentum space, with $\bfk_N$ the momentum of the nucleus. The effective axion-nucleus coupling is defined as
\begin{equation}\label{eq:gNdefinition}
    g_N \equiv \lambda_p g_p + \lambda_n g_n.
\end{equation}
with 
\begin{align}
    \lambda_p & = \frac{\langle  \sum_i^A {S}^z_{p_i} \rangle}{J}\label{eq:deflambdap}\\
    \lambda_n & = \frac{\langle \sum_i^{A-Z} {S}^z_{n_i} \rangle}{J}\label{eq:deflambdan}
\end{align}
The $\lambda_{p,n}$ parameters can be determined with the odd group model~\cite{Engel:1989ix} or with more modern nuclear wave functions calculations~\cite{Hu:2021awl}, as summarised in Tab.~\ref{tab:lambda}.

Since the crystal will in general contain multiple elements and multiple isotopes, we will need to index the effective axion-nucleus coupling with the lattice cell coordinate $\bfl$ and the position within the cell $j$. We further find it convenient to factor out the axion-proton coupling, by defining
\begin{equation}
\lambda_{\bfl j}\equiv g_{\bfl j}/g_p
\end{equation}
with $g_{\bfl j}$ defined by~\eqref{eq:gNdefinition}, applied to the atom at position $\bfl j$. In this manner the parameter $\lambda_{\bfl j}$ only depends on the UV structure of the axion model, through the $g_p/g_n$ ratio, and the nuclear matrix elements, through~\eqref{eq:deflambdap} and~\eqref{eq:deflambdan}. The $g_p$ will control the overall rate.

\begin{table}
\begin{tabular}{l c c c c c} 
  & nat.~ab. &  \; $J$ & \; $\lambda_{p}$ \; \; & $\lambda_n$  &ref\\ 
 \hline 
 $^{1}$H & 100\% & 1/2& 1.00 &  0 & /\\
 $^2$D & 0.01\% &1& 0.46  & 0.46 & \cite{1987inp..book.....K}\\
 $^7$Li & 92\% &3/2& 0.25  & 0 & \cite{Engel:1989ix} \\
 $^9$Be & 100\%&3/2 & 0 & 0.21 & \cite{Engel:1989ix}\\
 $^{27}$Al & 100\%&5/2& 0.11  & 0.014 & \cite{Hu:2021awl}\\
 $^{69}$Ga & 60\% & 3/2 & 0.07 & 0 & \cite{Engel:1989ix}\\
 $^{71}$Ga & 40\% & 3/2 & 0.15 & 0 & \cite{Engel:1989ix}\\
$^{75}$As & 100\% & 3/2 & \num{-0.007} & 0 & \cite{Engel:1989ix}
 \end{tabular}
 \caption{A list of the isotopes contained in the materials we consider, along with their natural abundance, spin, and $\lambda_{u,d}$. We use the shell model calculations in~\cite{Hu:2021awl} when available, and the odd group model calculations otherwise~\cite{Engel:1989ix}.}
 \label{tab:lambda}
 \end{table}

\subsection{P \& T violating forces for large nuclei\label{app:PTviolatingpot}}
In the heavy-pion limit
and nonrelativistic approximation, the potential experienced by the outer nucleon interacting with the nucleons in the core can be approximated by:
\begin{equation}
    V_{PT} = \frac{1}{2 m_i m_\pi^2} \eta_i \Bsigma_N\cdot \mathbf{\nabla}\rho(\mathbf{r})
\end{equation}
where $\rho(\mathbf{r})$ is the number density of core nucleons, normalised as $\int d^3 r\, \rho(r) = A$, and
\begin{equation}
    \eta_p = -\eta_n \approx g_s\left(\frac{g_0}{5}-g_1\right)
\end{equation}
where the index $i$ refers here to the species of nucleon, e.g. either a proton or neutron, $g_s$ is the standard P and T preserving pion-nucleon coupling, and $g_0$ and $g_1$ denote the isoscalar and isovector P and T violating pion-nucleon couplings, with values~\cite{deVries:2020iea}
\begin{align}
    g_s &\approx -13.45 \\
    g_0 &\approx (15.5 \pm 2.5) \times 10^{-3} \frac{a(t)}{f_a} \\
    g_1 &\approx -0.2\,g_0.
\end{align}
Using the estimate demonstrated in chapter 10 of~\cite{Khriplovich:1997ga},  for $A \gg 1$, the above potential generates a nuclear EDM through a polarisation of the nuclear charge distribution, with approximate size and scaling:
\begin{equation}
    d_{(2)} \approx -e\left(q - \frac{Z}{A}\right)\,\zeta_i\, \frac{1/2 - K}{J+1}
\end{equation}
where $q=1$ or $0$ depending on whether the valence nucleon is a proton or neutron respectively, $K=\frac{l-J}{2J-1}$ and $l$ is the nuclear orbital angular momentum. The coefficient $\zeta$ contains all the information about the pion-axion induced potential:
\begin{equation}
    \zeta_i \simeq \frac{15}{2 \pi^2 m_i}\, \eta_i.
\end{equation}
We therefore find the nuclear EDM in large A nuclei is roughly of size:
\begin{equation}
    d_{(2)} \sim  e\left(q-\frac{Z}{A}\right) \frac{a(t)}{f_a} \frac{1}{\pi^2 m_i} \sim 10^{-15}\,\frac{a(t)}{f_a}\,\rm{e}\,\rm{cm}.
    \end{equation}

\section{The Born-Oppenheimer (or adiabatic) Approximation}
\label{app:bo}

In this appendix, we will demonstrate how the electronic degrees of freedom decouple from the nuclear center of mass degrees of freedom for the operators and energy scales considered in this work. The derivation presented here is based on Ref.~\cite{broglia2004}. 

We take the non-relativistic total Hamiltonian of the crystal to be
\begin{align}
    H_{tot} &= H_0 + \delta H_a(t) \\
    H_0 &= -\sum_n \frac{\nabla_n^2}{2\,m_n} - \sum_i \frac{\nabla_i^2}{2\, m_e} + V_{nuc,nuc} + V_{e,e} + V_{nuc,e}
\end{align}
where $n$ denotes a nuclear coordinate, an $i$ is an electron coordinate. We denote the set of the nuclear center of mass positions by $\{\mathbf{R}\}$ (we treat the nuclei as rigid bodies as we work at energies far below the internal nuclear excitation energies) and electron positions by $\{\mathbf{r}\}$. The various potentials are
\begin{equation}
    V_{nuc,nuc}\left(\{\mathbf{R}\}\right) = \frac{1}{2}\sum_{m,n\neq m}\frac{Z_n Z_m e^2}{\lvert \mathbf{R}_n- \mathbf{R}_m \rvert}
\end{equation}
\begin{equation}
    V_{e,e}\left(\{\mathbf{r}\}\right) = \frac{1}{2}\sum_{i,j\neq i}\frac{e^2}{\lvert \mathbf{r}_i- \mathbf{r}_j \rvert}
\end{equation}
\begin{equation}
    V_{nuc,e}\left(\{\mathbf{r}\}, \{\mathbf{R}\}\right) = \frac{1}{2}\sum_{i}\sum_n\frac{Z_n e^2}{\lvert \mathbf{R}_n- \mathbf{r}_i \rvert}
\end{equation}
The axion generates two time-dependent perturbations through its coupling to the nucleus:
\begin{equation}
    \delta H_a(t) = \delta H_{EDM}(t) + \delta H_{ann}(t)
\end{equation}
Due to Schiff's theorem~\cite{schiff1963}, the EDM operator $\delta H_{EDM}(t)$ has vanishing diagonal matrix elements (i.e. no energy shift) in the limit of point-like, non-relativistic nuclei. This is famously violated by using large, heavy nuclei, which are irrelevant to this work. Likewise, the ``axioelectric" axion-nucleon coupling $\delta H_{ann}(t)$ also has no diagonal matrix elements in the limit of light nuclei (the violation by heavy nuclei arises from spin-orbit effects~\cite{Berlin:2023ubt}).

As shown in~\cref{sec:EDMsection}, the matrix elements induced by the EDM are directly proportional to those induced by the nucleon-axion coupling. Therefore, we here show why the (off-diagonal) matrix elements of the latter can be computed while ignoring the electronic degrees of freedom, as done in the main text, and one can easily use the same arguments for the former.

We may write the full spatial wavefunction of the system with quantum numbers $Q$, $\Phi_Q$, defined by $H_0\Phi_Q=E_Q\Phi_Q$ as 
\begin{equation}\label{eq:fullfunc}
\Phi_{Q}\left(\{\mathbf{r}_i\}, \{\mathbf{R}\}\right)=\sum_{m}\psi_{Q,m}\left(\{\mathbf{R}\}\right)\phi_{e,m}\left(\{\mathbf{r}\}, \{\mathbf{R}\}\right),
\end{equation}
where $\psi_{Q,m}$ is the part of the wave function describing the degrees of freedom of the nuclei.
The $\phi_{e,m}$ are the eigenstates of the electronic Hamiltonian with the nuclei kept at fixed positions $\{\bm{R}\}$:
\begin{equation}
     \left( \sum_i -\frac{\nabla_i^2}{2 m_e} +V_{nuc,e} +V_{e,e} \right) \phi_m = \mathcal{E}_m(\{ \mathbf{R}\}) \phi_{e,m},
\label{eq:psi_eigenvalues}
\end{equation} 
with $\mathcal{E}_m( \{\mathbf{R}\})$ as the nuclear-location-dependent eigen-energies. The difference between two electronic energy levels (for the same fixed nuclear locations) is therefore going to be of the order of the electronic energy levels, which are far above the phonon energy levels in the systems considered.

Using the orthonormality of the $\phi_{e,m}$ and~\eqref{eq:psi_eigenvalues}, one can show that~\cite{broglia2004}
\begin{align}
    &\sum_n \left(\frac{-\nabla_n^2}{2m_n} + V_{nuc,nuc} + \mathcal{E}_m(\{ \mathbf{R}\}) \right)\psi_{Q,m}\nonumber\\
    &+ \sum_{m'} C(m,m')\, \psi_{Q,m'} = E_Q \,\psi_{Q,m},
\end{align}
where we have defined
\begin{align}
    C(m,m') =&  -\frac{1}{2m_n} \sum_n 2 \bra{\phi_{e,m}} \boldsymbol{\nabla}_n   \ket{\phi_{e,m'}} \boldsymbol{\nabla}_n\nonumber\\ 
    &+ \bra{\phi_{e,m}}\nabla_n^2\ket{\phi_{e,m'}}.
    \label{eq:BOapprox}
\end{align}
Here, for an operator $\mathcal{O}$, the expression $\bra{\phi_{e,m}}\mathcal{O}\ket{\phi_{e,m'}}$ is meant to convey an integration only over the electronic degrees of freedom, $\{{\bf r}\}$, and so $C(m,m')$ also depends on $\{{\bf R}\}$.

The Born-Oppenheimer, or adiabatic, approximation amounts to neglecting the terms $C(m,m')$, which correspond to the effect of the nuclear motion on the waveform of the electron. Corrections to this approximation are assumed to scale as $m_e/m_n$. This approximation has been widely tested experimentally and computationally (see e.g. \cite{B211193D}). Neglecting it, the $\psi_{Q,m}$ now obey a Schrodinger equation with an effective $\{{\bf R}\}$-dependent potential. Furthermore, since in the absence of $C(m,m')$, the different $m$ don't mix in~\eqref{eq:BOapprox}, we may drop the $m$ sum in~\eqref{eq:fullfunc}.  It is the $\psi$, rather than $\Phi$ that we use in the main text, and they describe the wavefunctions of the atomic degrees of freedom.

We can now examine the axion-induced operators. As previously mentioned, we will focus on $g_{ann}$

\begin{equation}
\begin{split}
\delta H_a(t) \Phi_{Q} = \sum_{n}  g_{ann} &\left(\partial_t a \right) \left(\boldsymbol{\sigma}_n \cdot \left(\frac{-i \boldsymbol{\nabla}_n}{m_n}\phi_{e,m}\right.\right)\psi_{Q,m}
 \\ &\left.+ \phi_{e,m}\boldsymbol{\sigma}_n 
 \cdot \frac{-i \boldsymbol{\nabla}_n}{m_n}\psi_{Q,m}\right)
\end{split}
\end{equation}
From~\eqref{eq:BOapprox}, we see that the operator here $\nabla_n$ applied on the electronic degrees of freedom is what we neglect in the Born-Oppenheimer approximation. In order to get a complete decoupling of the axion operator from the electronic degrees of freedom, we must make a further assumption that the nuclear spin operator does not act on $\phi_{e,m}$. The exact conditions for this to occur are beyond the scope of this work, however, a sufficient condition fulfilled by all the materials we examine is that there are no unpaired electrons. Therefore we find
\begin{equation}
    \delta H_a(t) \Phi_{Q,m} = \phi_{e,m} \left( \delta H_a(t) \,\psi_{Q,m}\right)
\end{equation}
and we see that the axion perturbation does not affect the electronic degrees of freedom of the system. This should remain valid as long as the frequency associated with the perturbations is much lower than the difference in electron energy levels, $\omega_a \ll \Delta \mathcal{E}_m$.

\section{Further details on coherent absorption}
\label{app:coherent}
In this appendix, we give further details on the derivation of the coherent absorption rate of axions for diatomic crystals. Since the coherent process only excites approximately zero-momentum phonons, these must be optical phonons in order to have non-vanishing energy.
We also elaborate on the averaging over different atomic isotopes in this scenario.
\subsection{Diatomic rate}

To evaluate the rate, we need the polarisation vectors of the optical phonons. For diatomic crystals, these are known at long wavelengths to be \cite{Kittel2004}
\begin{align}
    \Bepsilon_{1, \mathrm{LO} ,\mathbf{k}} &\approx \sqrt{\frac{A_1}{\sum_{j '} A_{j '}}} \hat{\mathbf{k}}, \nonumber
    \\
    \Bepsilon_{2 ,\mathrm{LO} ,\mathbf{k}} &\approx - \sqrt{\frac{A_2}{\sum_{j '} A_{j '}}}  e^{-i \mathbf{k} \cdot \mathbf{x}_2^0} \, \hat{\mathbf{k}},
\end{align}
for the longitudinal optical (LO) modes\footnote{We are assuming here and in the following that $2 \pi /L < k_a$, so that there is a state in the reciprocal lattice with small but non-vanishing momentum, so that the concept of longitudinal and transverse modes is well-defined.}, where $\mathbf{r}_2^0$ is the displacement of the second atom from the first atom. There are also two transverse optical (TO) modes, whose polarisation vectors are related to the TO modes
\begin{align}
    \Bepsilon_{j, \mathrm{TO}, \mathbf{k}} = \hat{\mathbf{n}} \times   \Bepsilon_{j, \mathrm{LO}, \mathbf{k}}
\end{align}
for two independent unit vectors $\hat{\mathbf{n}}$ normal to the phonon momentum $\mathbf{k}$. 

In the case of an LO phonon with $|\mathbf{k} \cdot \mathbf{r}_2^0| \ll 1$, we find
\begin{align}
   \Big| \sum_j \frac{\lambda_j J_j \epsilon^z_{j, LO, \mathbf{k}}}{\sqrt{m_j}} \Big|^2 &= \frac{(\lambda_1 J_1 - \lambda_2 J_2)^2}{m_1 + m_2} (\hat{\mathbf{z}}\cdot\hat{\mathbf{k}})^2 \\ 
    &= \frac{(\lambda_1 J_1 - \lambda_2 J_2)^2}{3(m_1 + m_2)},
\end{align}
where in the second line we have averaged over the axion momentum, taken to be isotropic, so that $\langle (\hat{\mathbf{z}} \cdot \hat{\mathbf{k}})^2 \rangle = \langle (\hat{\mathbf{z}} \cdot \hat{\mathbf{k}}_a)^2 \rangle = 1/3$ by momentum conservation. In reality, the axion momentum would exhibit daily anisotropy due to the Earth's motion through the DM halo, although we ignore this effect here. Using this result and dividing by the total mass, we find~\eqref{eqn:coherentRate} for the rate per unit mass. The case of the two TO phonons follows in the same vein, with $\langle (\hat{\mathbf{z}} \cdot (\hat{\mathbf{k}} \times \hat{\mathbf{n}} )^2 \rangle = 1/3$.

\subsection{Isotopic averaging}
\label{app:isotopes}

In writing~\eqref{eq:ratebeforefork}, we are implicitly assuming that we have isotopically pure materials. If there are multiple isotopes present in a material, then we must take this into account in the mass and nuclear form factor of a lattice site. In principle, also the periodicity of the lattice could be broken, however, since we work in crystals that are either dominated by a single isotope, or all isotopes have nearly identical mass, we may assume that $D_j(\omega)$ is isotope independent. In such scenarios, the sum over initial configurations is only slightly more complicated, by containing a sum over a homogeneous and isotropic ensemble of isotope distributions, labeled by $i'$. If the spin state $|s^i_{\bfl j}\rangle$ is independent of the isotope (as is the case for $^{69}$Ga and $^{71}$Ga), then the effect of this averaging is
\begin{align}
    \sum_{i'} w_{i'} \frac{\lambda^{i'}_{\bfl j} \lambda^{i'}_{\bfl' j'}}{\sqrt{m_{\bfl j} m_{\bfl' j'}}} = \frac{\bar{\lambda}_{\bfl j} \bar{\lambda}_{\bfl' j'}}{\sqrt{m_{\bfl j} m_{\bfl' j'}}},
\end{align}
where $\bar{\lambda}_{\bfl j}$ is the form factor of a site averaged over the natural abundance, and we have neglected the subleading isotope dependence of the mass.

\section{Angular average in incoherent absorption\label{app:angularaverage}}

In this appendix, we derive~\eqref{eqn:ensembleAverage} for the effective coupling $\xi_j$ between an axion and a phonon in a crystal with a random, isotropic ensemble of nuclear spins. The ensemble-averaged matrix element we require is \begin{equation}
 \mathcal{M}^{\alpha \beta}_j = \frac{1}{N}\sum_{\bfl}\sum_{i} w_i  \frac{\lambda_{\bfl j}^2}{m_{\bfl j}} \langle s^i_{\bfl j} |J^\alpha_{\bfl j} J^\beta_{\bfl j}|s^i_{\bfl j}\rangle,
\end{equation}
where the weights $w_i$ satisfy \begin{align}
    \sum_i w_i = 1.
    \label{eqn:weightNorm}
\end{align}

Since the ensemble of crystal spins is isotropic, $\mathcal{M}^{\alpha \beta}_j$ must be proportional to the unit tensor, and so
\begin{align}
  \sum_i w_i  \langle s^i_{\bfl j} |J^\alpha_{\bfl j} J^\beta_{\bfl j}|s^i_{\bfl j}\rangle = c_{\bfl j} \delta^{\alpha \beta},
\end{align}
where $c_{\bfl j}$ is a constant. Taking the trace of this equation and using the normalisation~\eqref{eqn:weightNorm}, we find
\begin{align}
    c_{\bfl j} = \frac{1}{3} J_{\bfl j}(J_{\bfl j} + 1).
\end{align}
Putting this all together, we see
\begin{align}
    \mathcal{M}_j^{\alpha \beta} = \frac{1}{3N} \sum_{\bfl} \frac{\lambda_{\bfl j}^2}{m_{\bfl j}} J_{\bfl j}(J_{\bfl j} + 1) \, \delta^{\alpha \beta},
\end{align}
or 
\begin{align}
    \xi_j = \frac{1}{N}\sum_{\bfl} \frac{\lambda_{\bfl j}^2}{m_{\bfl j}} J_{\bfl j}(J_{\bfl j} + 1),
\end{align}
which is a crystal-lattice averaged coupling to the $j$th atom in the unit cell. 

\end{document}